# Transport Spectroscopy of Symmetry-Broken Insulating States in Bilayer Graphene

J. Velasco Jr., L. Jing, W. Bao, Y. Lee, P. Kratz, V. Aji, M. Bockrath, C.N. Lau[*] and C. Varma

*Department of Physics and Astronomy, University of California, Riverside, CA 92521*

R. Stillwell and D. Smirnov

*National High Magnetic Field Laboratory, Tallahassee, FL 32310*

Fan Zhang, J. Jung and A.H. MacDonald

*Department of Physic, University of Texas at Austin, Austin, TX 78712*

**The flat bands in bilayer graphene(BLG) are sensitive to electric fields $E_\perp$ directed between the layers, and magnify the electron-electron interaction effects, thus making BLG an attractive platform for new two-dimensional (2D) electron physics[1-5]. Theories[6-16] have suggested the possibility of a variety of interesting broken symmetry states, some characterized by spontaneous mass gaps, when the electron-density is at the carrier neutrality point (CNP). The theoretically proposed gaps[6,7,10] in bilayer graphene are analogous[17,18] to the masses generated by broken symmetries in particle physics and give rise to large momentum-space Berry curvatures[8,19] accompanied by spontaneous quantum Hall effects[7-9]. Though recent experiments[20-23] have provided convincing evidence of strong electronic correlations near the CNP in BLG, the presence of gaps is difficult to establish because of the lack of direct spectroscopic measurements. Here we present transport**

[*] E-mail: lau@physics.ucr.edu

**measurements in ultra-clean double-gated BLG, using source-drain bias as a spectroscopic tool to resolve a gap of ~2 meV at the CNP. The gap can be closed by an electric field $E_\perp \sim$ 13 mV/nm but increases monotonically with a magnetic field $B$, with an apparent particle-hole asymmetry above the gap, thus providing the first mapping of the ground states in BLG.**

The single-particle band structure of BLG resembles that of a gapless semi-conductor, with parabolic valence and conduction bands touching at the highly symmetric $K$ and $K'$ Dirac points. When weak remote hopping processes are included, the momentum space band-touching point splits in four[4] and Liftshitz transitions occur at low carrier densities as discussed in Supplementary Information[24]. For $E_\perp \neq 0$, a band gap develops and increases with $E_\perp$, saturating at ~0.3 eV,[4,25,26] and the conduction band has a "Mexican hat" shape. When electron-electron interactions are included BLG is expected to be unstable to broken-symmetry states that can be viewed as layer-pseudospin ferromagnets.[6]

BLG has been a discussed using a two-band model[8] valid near the CNP in which the broken-symmetry state quasiparticle Hamiltonian

$$H = -\left(\frac{p^2}{2m^*}\right)\left[\cos(2\phi_\mathbf{p})\sigma_x \pm \sin(2\phi_\mathbf{p})\sigma_y\right] - \Delta \cdot \boldsymbol{\sigma} \ . \qquad (1)$$

The first term in Eq.(1) is the quadratic band Hamiltonian, where $\tan\phi_\mathbf{p}=p_y/p_x$, $\mathbf{p}$ is the angular momentum, $m^*$ is the effective mass of the carriers, $\sigma$ is a Pauli matrix vector that acts on the layer degree of freedom, $\Delta$ represents the order parameter of the broken symmetry state, and the $\pm$ signs refers to the K and K' valleys, respectively. It has been variously predicted[6-10] to be oriented in the $\pm z$ direction, yielding gapped isotropic states with large momentum-space Berry curvature[8], or in the *xy* plane, yielding gapless anisotropic (nematic) states[11,12] with vanishing Berry curvatures. A variety of distinct but related massive states occur, depending on the dependence of the sign of $\Delta_z$ on spin and valley.[8] The attributes of some theoretically proposed BLG states are summarized in Table 1 and Fig.1, and the relationship between $\Delta$ and quasiparticle electronic properties is discussed in detail in [24].

In the first experimental reports[20,21], which hinted at an ordered state in BLG, the minimum conductivity $\sigma_{min}$ was a non-monotonic function of $E_\perp$, with a value ~60 µS at $n=E_\perp=0$. The experimental evidence for spontaneous gaps in this work was not conclusive, however, and the nature of the ground state at the CNP has remained controversial. We demonstrate here that the ground state is indeed gapful with a magnetic field (*B*) -dependent gap of the form

$$E_{gap} = \Delta_0 + \sqrt{a^2 B^2 + \Delta_0^2} \qquad (2)$$

with $|\Delta_0|$ ~1 meV and $a$ ~ 5.5 meV/T. By tracking the dependence of $E_{gap}$ on $B$ and on $E_\perp$, which is believed to induce a transition between a layer unpolarized and a layer polarized state, we are able to provide the first mapping of BLG ground states.

Our devices consist of exfoliated BLG sheets with Cr/Au electrodes suspended between Si/SiO$_2$ back gates and metal top gates (Fig. 1a), with very high mobility (~80,000 – 100,000 cm$^2$/Vs ). By tuning voltages applied to the back gate $V_{bg}$ and top gate $V_{tg}$, we can independently control $E_\perp$ and the charge density $n$ induced in the bilayer. Fig. 1c plots the two-terminal differential conductance $G=dI/dV$ at $V=0$ of the device (color) *vs.* $V_{bg}$ and $B$ and reveals Landau levels as colored bands radiating from the CNP and $B=0$. Line traces of $G(V_{bg})$ at constant $B$ exhibit conductance plateaus with values near 0, 1, 2, 3, 4 and 8 $e^2/h$ (Fig. 1d), indicating that the

8-fold degeneracy of the lowest Landau level (LL)[27-30] is broken. (Here $e$ is the electron charge and $h$ the Planck constant.) Observation of these plateaus at relatively low $B$ underscores the high quality of the device.

Close inspection of Fig. 1d reveals that the $v=0$ gap appears to persist down to $B=0$. Indeed, at $B=0$, the $G(E_\perp,n)$ plot shows a *local minimum* at $n=E_\perp=0$ (Fig. 2a-b). This contradicts the single-particle picture, which predicts a gap that is roughly linear in $E_\perp$, hence a monotonically decreasing $G(|E_\perp|)$. Our data therefore suggests a breakdown of the non-interacting electron picture, even allowing for the possibility[24] of uncontrolled mechanical deformations in our suspended flakes[31].

To investigate the CNP resistive state spectroscopically, we measure $G$ as a function of source-drain bias $V$ and $E_\perp$ while keeping $n=0$. Typical spectroscopic transport measurements are performed using tunnel probes, whereas our devices have highly transparent contacts. Because they are in the quasi-ballistic limit, we nevertheless anticipate spectroscopic resolution. Our non-linear transport data is summarized in Fig. 2c-f. The most striking feature is the region of dark blue/purple at the center of the plots, corresponding to the highly resistive state at small $E_\perp$. At $n=E_\perp=V=0$, the device is *insulating*, with $G_{min}<\sim0.5$ µS (Fig 2f, green curve). We note that this insulating state is observed only in high mobility samples realized after current annealing[32], and then only at low sweeping rates[24] and after careful optimization of the measurement setup to eliminate spurious voltage noise. As $V$ increases, $G$ remains approximately 0 until it increases abruptly at $V\sim\pm1.9$ mV, reaching sharp peaks before it decreases again to $\sim300$ µS. The $G(V)$ curve bears a striking resemblance to the BCS superconducting density of states, and strongly suggests the formation of an ordered phase with an energy gap $E_{gap}\sim1.9$ meV. Importantly, this gap can be closed by application of $E_\perp$ of *either* polarity: $G$ increases with $|E_\perp|$ (Fig. 2d); upon application of moderate $E_\perp > \sim 13$ mV/nm, the BCS-like structure completely vanishes and the $G(V)$ curve becomes approximately V-shaped, with a finite conductance minimum of ~100 µS at $n=0$ (Fig. 2f, purple curve). Finally, for sufficiently large $E_\perp$, $G(V=0)$ start to decrease with increasing $E_\perp$ (Fig. 2d), reverting to single-particle behavior.

Such an insulating state, which is the most salient experimental feature, is only observed in devices with the highest mobility. To gain further insight, we study its evolution with $B$ (Fig. 3). When $B$ increases from 0, the gapped phase continuously evolves into a bilayer insulating state at filling factor $\nu=0$, with the sharp peaks in $G$ becoming sharper and more dramatic. The magnitude of the gap, as measured from the bias values of the sharp peaks in $G$, is well-described by Eq. (2), where $|\Delta_0| \sim 1$ meV obtained from Fig. 2f, and $a=5.5$ mV/T. For $B>0.5$T, $E_{gap} \sim 5.5\ B$ (meV/T) which is much larger the single-particle gap induced by Zeeman splitting, ~0.1 meV/T. We note that our observation of a linear dependence of $\Delta_{\nu=0}$ on $B$ is consistent with previous reports[7,29], but with a significantly larger magnitude, possibly due to the superior quality of our device. Another noteworthy feature of Fig. 3a is that the conductance peaks at positive and negative bias voltages, which we associate with conduction and valence band edges, are highly asymmetric. The asymmetry increases with increasing $B$, and reverses when $B$ changes sign, suggesting particle-hole asymmetry in the device.

To summarize our experimental findings: (A) ultra-clean BLG is insulating at $n=B=E_\perp=0$, with an energy gap $E_{gap} \sim 1.9$ meV that can be closed by $E_\perp$ of either sign; (B) the energy gap evolves in $B$ following Eq. (2); and (C) this state is apparently particle-hole asymmetric. These observations provide much insight into the nature of BLG's symmetry-broken ground state. For instance, observation (A) rules out gapless ordered states[11,12]. For the gapped states, the symmetric dependence on $E_\perp$ indicates that there is no net charge imbalance between the two layers, thus excluding states with net spontaneous layer polarization like the QVH state depicted in Fig. 1. (We note that the QVH state is expected to be the ground state under sufficiently large $E_\perp$, as observed experimentally.)

Thus, among the proposed states, we are left with the 3 gapped state candidates that have no overall layer polarization. Given the flavor (spin-valley) symmetries, electrons in each layer can be valley polarized and form a quantum anomalous Hall insulator (QAH)[7,8,33,34], spin polarized to form a layer antiferromagnet (LAF)[8], or neither to form a quantum spin Hall (QSH) insulator[7,8,35-37] (Fig. 1b). Mean-field calculations indicate that these states are comparable[9] in energy and that the QVH is stabilized by electric fields $E_\perp$ ~ 5-20 mV/nm[7,9], in agreement with the critical field value of ~ 13 mV/nm observed experimentally. For all three states, Eq. (1) predicts a gap within each valley with a $B$-dependence described by Eq. (2), with parameters $\Delta_z = \Delta_0$ and $a = \sqrt{2\hbar e/m^*}$. From $a$=5.5 meV/T, we obtain $m^*$=0.03 $m_e$ where $m_e$ is electron rest mass, in good agreement with other measurements[38,39] of the interaction enhanced[40,41] quasi-particle mass near the BLG CNP. Thus our observations demonstrate unequivocally the symmetry-broken gapped phase in charge neutral graphene.

Nailing down the exact phase of the ground state, however, is considerably more difficult. Eq. (2) applies to charges in a single valley. If we assume ideal coupling of both layers to electrodes, none of the candidate states can account for all aspects of our data. The LAF state would have a $B$-independent gap. Both QAH and QSH states are ruled out by their topologically protected edge states which are expected to yield two-terminal conductances ~ $4e^2/h$ or 154 μS.

The QAH state is also ruled out because the density position of its gap is expected to deviate from the CNP in a finite magnetic field, following lines with $\nu=4$; yet it is the only state with particle-hole asymmetry and valley-indirect gaps that are in agreement with Eq. (2).

On the other hand, we note that since most of the metals are deposited on the top layer and contact the bottom layer only via the edges, the electrodes could couple preferentially to one layer (or equivalently, one valley). In this case, the LAF state, which is the only proposed insulating phase, can account for both observations (A) and (B). Since the absence of edge states is the most robust experimental signature, our observations are most consistent with the LAF state. We note that, however, the particle-hole asymmetry in $B$ is not explained within this picture. It is thus possible that a new ground state that has not been theoretically proposed underlie our observations. Further theoretical and experimental work will be necessary to ascertain the nature of the gapped state and achieve full understanding of our observations.

Lastly, we focus on the QH states in external electric and magnetic fields and further establish the bias-dependent measurement as a spectroscopic tool. Fig. 4a plots $G$ vs. $E_\perp$ and $n$ at constant $B=3.5T$. As shown by the line traces in Fig. 4b-c, at $E_\perp=0$, only the $\nu=0$ and $\nu=4$ QH plateaus are observed; at finite $E_\perp$, all integer QH states between 0 and 4 are resolved, demonstrating degeneracy-lifting by the perpendicular electric field. From the spectroscopic data $G(V,E_\perp)$ at constant $B$, the gap $\Delta_{\nu=0}$ is diamond-shaped (Fig. 4d-e): its magnitude decreases linearly with applied $E_\perp$ of either polarity, until it is completely closed at a critical field $E_\perp^*$. Fig. 4f plots $E_\perp^*$ obtained at 3 different $B$ values. The data points fall on a straight line, with a best-fit slope of ~ 12.7mV/nm/T[42]. Extrapolation of the best-fit line to $B=0$ yields a finite $E_\perp^*$ intercept ~ 12.5 mV/nm, which agrees with the critical $E_\perp$ value estimated from the zero $B$ field data in Fig.

2d. Both the slope and the finite $E_\perp^*$ intercept are consistent with those measured from the movement of the $\nu$=4 plateau as functions of $E_\perp$ and $B^{21,24}$. Taken together, our data indicate a transition from quantum Hall ferromagnet to the QVH state, and confirms that the bias-dependent measurement provides a spectroscopic determination of the $\nu$=0 gap.

**Acknowledgements** We thank Rahul Nandkishore, Ben Feldman, Amir Yacoby, Leonid Levitov, Pablo Jarillo-Herrero and Kostya Novoselov for stimulating discussions, and Dexter Humphrey, Gang Liu, Adam Zhao and Hang Zhang for assistance with fabrication. This work was supported in part by UC LabFees program, NSF CAREER DMR/0748910, NSF/1106358, ONR N00014-09-1-0724, and the FENA Focus Center. D.S. acknowledge the support by NHMFL UCGP #5068. Part of this work was performed at NHMFL that is supported by NSF/DMR-0654118, the State of Florida, and DOE. A.M., J. J. and F. Z. acknowledges the support by Welch Foundation grant TBF1473, NRI-SWAN, and DOE grant DE-FG03-02ER45958. C.V. acknowledges the support by NSF DMR-0906530. V.A. acknowledges the support by UCR I.C.

Table 1. Attributes of possible ordered states in BLG at $n=E_\perp=0$.

|  | Nematic Order | QAH | QSH | LAF | QVH |
|---|---|---|---|---|---|
| Gapped? | No | Yes | Yes | Yes | Yes |
| 2-terminal $\sigma_{min}$ | finite | $4e^2/h$ | $4e^2/h$ | 0 | 0 |
| Broken Symmetries | in-plane rotation | time reversal; Ising Valley | spin rotational; Ising Valley | time reversal; spin rotation | inversion |

Figure 1. **a,** SEM image and schematics of a device. **b,** Spin-valley configurations of the electrons in BLG for several possible phases. Here red (blue) arrows indicate electrons from the K (K') valley. **c,** Differential conductance $G=dI/dV$ vs. $B$ and $V_{bg}$ showing the Landau fan diagram. **d,** Line traces $G(V_{bg})$ along the dotted line in c at $B=1.3$, 2.8 and 6T, respectively.

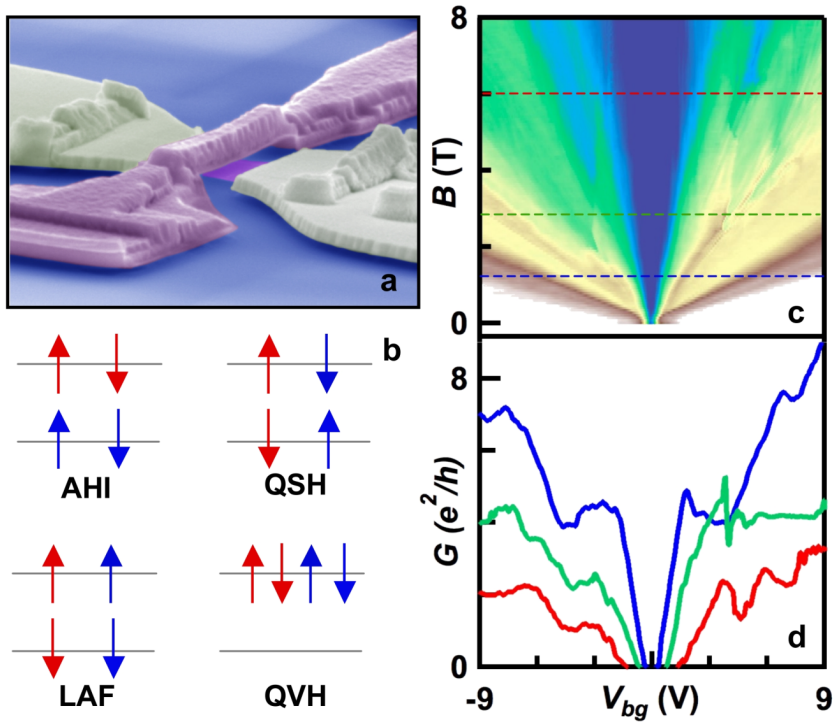

Figure 2. Transport data at $B=0$ and $T=300$mK. **a-b,** $G(n, E_\perp)$ and line traces $G(n)$ at $E_\perp=-37.5$, -25, -12.5, 0, 12.5, 25 and 37.5 mV/nm, respectively (left to right). The line traces are laterally offset for clarity. **c-d,** Large range $G(V, E_\perp)$ and line trace $G(E_\perp)$ at $n=0$. **e.** High resolution $G(V, E_\perp)$ for small bias range at $n=0$. **f,** Line traces $G(V)$ along the dotted lines in e at $E_\perp=0$ and -15 mV/nm, respectively.

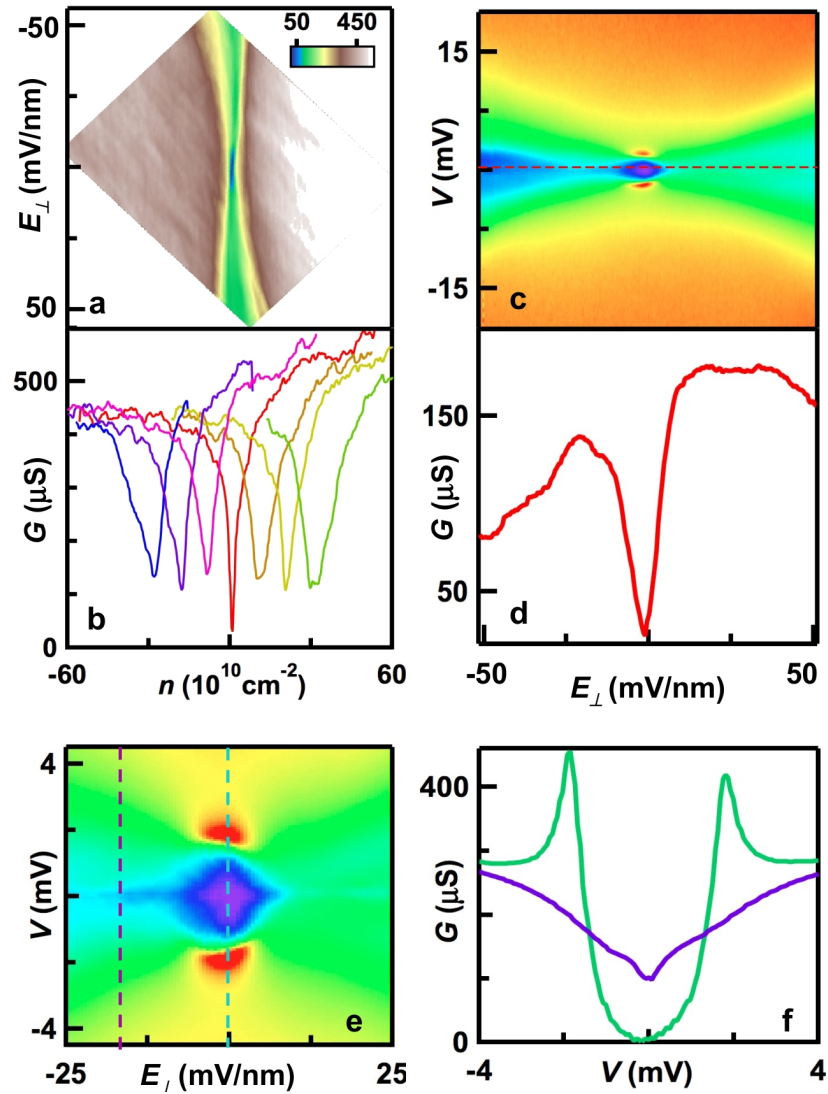

Figure 3. Transport data in magnetic field at *n=0* and $E_\perp=0$. **a-b,** *G(V, B)* and line traces *G(V)* at *B*=-0.4 (red), -0.2 (orange), 0 (green), 0.2 (blue) and 0.4 T (purple), respectively. **c,** Large range *G(V, B)*. The dotted line plots Eq. (2) with $\Delta_0$ =1 meV, and *a*=5.5 meV/T. **d,** Line traces *G(V)* at *B*=2 (red), 3 (orange), 4 (green), 5(cyan) and 6T (blue). Inset : Current-voltage characteristics at *B*=8T.

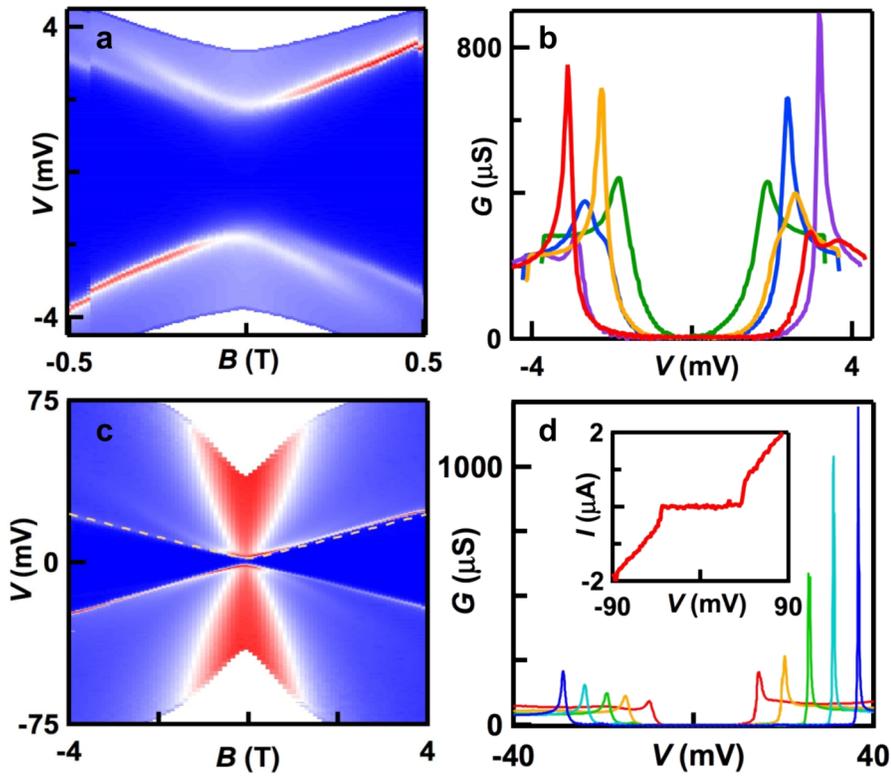

Figure 4. Transport data at constant B. **a,** $G(n, E)$ at $B=3.5$T. **b-c,** Line traces $G(n)$ along the dotted lines in a at $E_\perp=0$ and -20 mV/nm. **d-e,** $G(V, E)$ at $B=1.3$T and 3.6T, respectively. **f,** Critical electric field $E^*_\perp$ vs. B. The blue line is a best-fit to data, with a slope of 12.7 mV/nm/T.

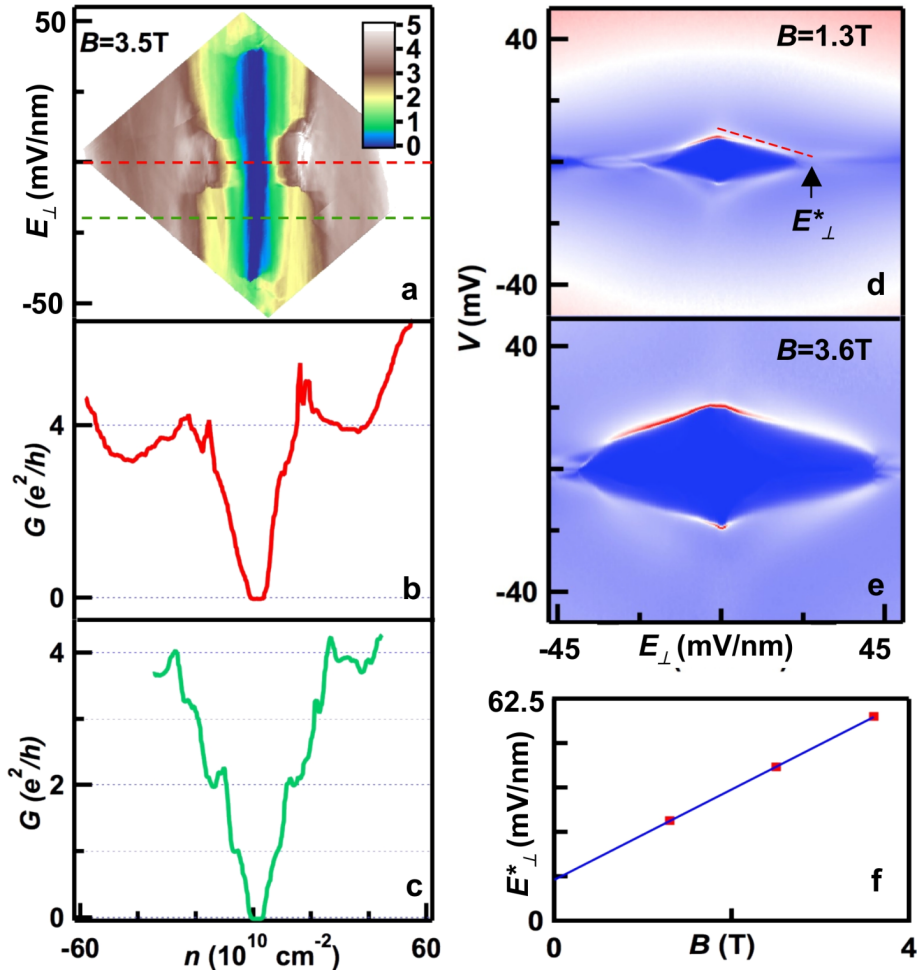